\documentclass{llncs}
\usepackage{graphicx}
\DeclareGraphicsExtensions{.pdf,.png,.jpg}

\usepackage{float}
\floatstyle{boxed} 
\restylefloat{figure}

\usepackage{subfig}

\usepackage{macros}

\newcommand\gate[1]{{\tt #1}}
\newcommand\circuit[2]{\mathbb{C}~#1~#2}
\newcommand\ser{ \rhd }
\newcommand\para{ \& }
\newcommand\invar[5]{{#3} \vdash_{#2}^{#1} #4 \bowtie #5}
\newcommand\unit{{\bf 1}}

\newcommand\bool{\mathbb{B}}
\newcommand\data{\mathbb{T}}
\newcommand\word[1]{\mathbb{W}_{#1}}
\newcommand\cword[1]{\mathbb{W}_{\tt #1}}
\newcommand\sumn[2]{{\tt sumn}~\unit_{\tt #1}~{\tt #2}}

\title{Coquet: a Coq library for verifying hardware}
\author{Thomas Braibant}
\institute{LIG, UMR 5217, INRIA}

\begin{document}
\maketitle{}
\begin{abstract}
  We propose a new library to model and verify hardware circuits in
  the Coq proof assistant.
  This library allows one to easily build circuits by following the
  usual pen-and-paper diagrams. We define a deep-embedding: we use a
  (dependently typed) data-type that models the architecture of
  circuits, and a meaning function.
  We propose tactics that ease the reasoning about the behavior of the
  circuits, and we demonstrate that our approach is practicable by
  proving the correctness of various circuits: a text-book divide and
  conquer adder of parametric size, some higher-order combinators of
  circuits, and some sequential circuits: a buffer, and a register.
\end{abstract}

\section*{Introduction}

Formal methods are widely used in the verification of circuit design,
and appear as a necessary alternative to test and simulation
techniques.
Among them, model checking methods 
have the advantage of being fully automated but can only deal with
circuit of fixed size and suffer from combinatorial explosion. On the
other hand, circuits can be formally specified and certified using
theorem provers~\cite{hanna-veritas,UCAM-CL-TR-77,DBLP:conf/msiw/HuntB89}.
For instance, the overall approach introduced
in~\cite{UCAM-CL-TR-77,melham} to model circuits in higher-order logic
is to use predicates of the logic to express the possible behaviour of
devices.



We present a study for specifying and verifying circuits in Coq. Our
motivations are two-fold.
First, there has been a lot of works describing and verifying circuits
in logic in the HOL and ACL2 family of theorem provers.  However, Coq
features dependent types that are more expressive. The Veritas
language experiment~\cite{hanna-veritas} hinted that these allow for
specifications that are both clearer and more concise. %
We also argue that dependent types are invaluable for developing
circuits reliably: some errors can be caught early, when type-checking
the circuits or their specifications.
Second, most of these works model circuits using a shallow-embedding:
circuits are defined as predicates or functions in the logic of the
theorem prover, with seldom, if any, way to reason about the devices
inside the logic: for instance, functions that operate on circuits
must be built at the meta-level~\cite{DBLP:journals/fac/SlindOIG07},
which precludes one from proving their correction. %
We define a data-type for circuits and a meaning function: we can
write (and reason about) Coq functions that operate on the structure
of circuits.

Circuit diagrams describe the wire connections between gates and have
nice algebraic properties~\cite{brown-hutton,lafont}. While we do not
prove algebraic laws, our library features a set of basic blocks and
combinators that allows one to describe such diagrams in a hierarchic
and modular way. We make precise the interconnection of circuits, yet,
we remain high-level because we make implicit the low-level diagram
constructs such as wires and ports.
Circuit diagrams are also used to present recursive or parametric
designs. We use Coq recursive definitions to \emph{generate} circuits
of parametric size, e.g., to generate a $n$-bit adder for a given
$n$. Then, we reason about these functions rather than on the tangible
(fixed-size) instantiations of such generators.
Circuits modelled by recursion have already been verified in other
settings~\cite{DBLP:conf/msiw/HuntB89,melham}.
The novelty of our approach is that we derive circuit designs in a
systematic manner: we structure circuits generators by mimicking the
usual circuit diagrams, using our combinators. Then, the properties of
these combinators allow us to prove the circuits correct.

We are interested in two kinds of formal dependability claims. First,
we want to capture some properties of well-formedness of the diagrams.
Second, we want to be able to express the \emph{functional
  correctness} of circuits -- the fact that a circuit complies to some
specification, or that it implements a given function.
Obviously, the well-formedness of a circuit is a prerequisite to its
functional correctness. We will show that using dependent types, we
can get this kind of verification for free. As an example, the
type-system of Coq will preclude the user to make invalid compositions
of circuits. 
%
Hence, we can focus on what is the intrinsic \emph{meaning} of a
circuit, and prove that the meaning of some circuits entails a
high-level specification, e.g., some functional program.

Our contributions can be summarized as follows: %
we propose a new framework to model and verify circuits in Coq
that allows to define circuits in a systematic
manner by following usual diagrams; %
we provide tactics that allow to reason about circuits; %
we demonstrate that our approach is practicable on practical examples:
text-book $n$-bit adders, high-level combinators, and sequential
circuits.

\subsubsection*{Outline.}
In \S\ref{sec:basics}, we give a small
overview of all the basic concepts underlying our methodology to
present how the various pieces fit together. We present the actual
definitions we use in \S\ref{sec:definitions}.  Then, in
\S\ref{sec:comb} and \S\ref{sec:time}, we demonstrate the feasibility
of our approach on some examples. We analyse some benefits of using a
deep-embedding in \S\ref{sec:free}. Finally, we compare our study to
other related work in \S\ref{sec:related-works}.

\section{Overview of our system}\label{sec:basics}
\begin{figure}[t]
  \centering
      \includegraphics*[height=4cm]{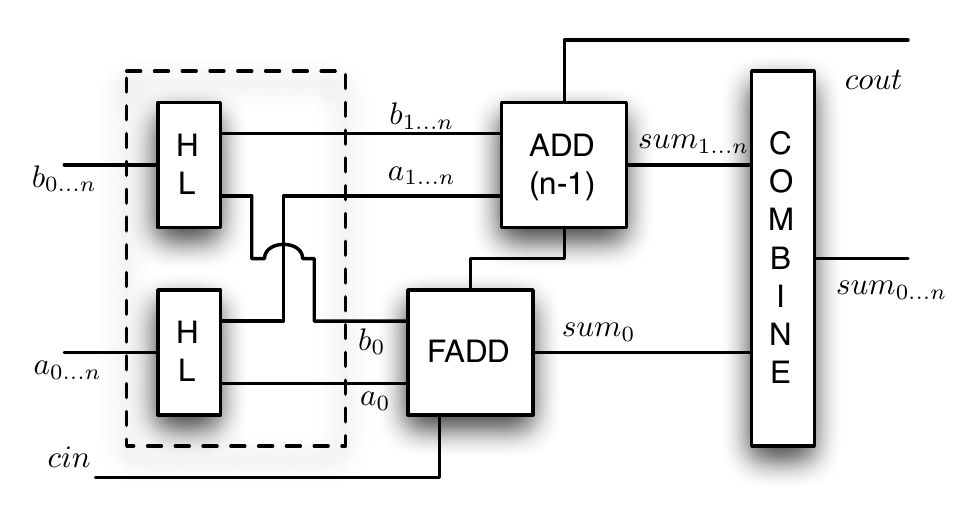}
  \caption{A recursive $n$-bit ripple-carry adder}
  \label{fig:intro-ripple}
\end{figure}

We give a global overview of the basic concepts of our methodology
first, before giving a formal Coq definition to these notions in the
next section. We take this opportunity to illustrate the use of our
system to represent parametrized systems through the example of a
simple $n$-bit adder: it computes an $n$-bit sum and a $1$-bit carry
out from two $n$-bit inputs and a $1$-bit carry in. The recursive
construction scheme of this adder is presented in
Fig.~\ref{fig:intro-ripple} (data flows from left to right), using a
\emph{full-adder}, i.e., a $1$-bit adder, as basic building
block. \thomas{hum, et les autres ?}

\subsubsection*{Circuit interfaces.}

Informally, we want to build circuits that relate $n$ input wires to
$m$ output wires, where $n,m$ are integers. For instance, the door
\gate{AND} has two inputs and one output.
However, using integers to number the wires does not give much
structure: the $n$-bit adder has $2n+1$ input wires, this does not
specify how they are grouped. 
Hence, we use arbitrary \emph{finite-types} as indexes for the wires
rather than using integers~\cite{harrison-euclidean}. A circuit that
relates inputs indexed by $n$ to outputs indexed by $m$ has type
$\circuit{n}{m}$, where $n$ and $m$ are types.  For instance, the
full-adder, a circuit with three inputs and one output, has type
\makebox{$\circuit{(\unit \oplus (\unit \oplus \unit))}{(\unit \oplus \unit)}$}, where $\oplus$ is the disjoint sum (associative to the
left) and $\unit$ is a singleton type. Hence, the $n$-bit adder has
type \makebox{$\circuit{(\unit \oplus \sumn{}{n} \oplus
    \sumn{}{n})}{(\sumn{}{n} \oplus \unit)}$}, where ${\tt sumn}~A~n$
is a $n$-ary disjoint sum.

\subsubsection*{Circuits combinators.}
The $n$-bit adder is made of several sub-components that are composed
together. We use \emph{circuit combinators} (or combining
forms~\cite{sheeran-muFP}) to specify the connection layout of
circuits.  
For instance, in Fig.~\ref{fig:intro-ripple}, the dashed-box is built
by composing in parallel two \gate{HL} circuits, that are then
composed serially with a combinator that reorders the wires. 
These combinators leave implicit the connection points in the
circuits, and focus on how informations flow through the circuit: the
wire names given in Fig.~\ref{fig:intro-ripple} do not correspond to
variables, and are provided for the sake of readability.

In our ``nameless'' setting, wires have to be forked and reordered
using \emph{plugs}: a plug is a circuit of type $\circuit{n}{m}$,
defined using a map from $m$ to $n$ that defines how to connect an
output wire (indexed by $m$) to an input wire (indexed by $n$).
Since we use functions rather than relations, this definition naturally
forbids short-circuits (two input wires connected to the same output
wire).

\subsubsection*{Meaning of a circuit.}
We now depart from the syntactic definitions of circuits to give an
overview of their semantics. We assume a type $\data$ of what is
carried by the wires, for instance booleans ($\bool$) or streams of
booleans (${\tt nat} \rightarrow \bool$).
Let $x$ be a circuit of type $\circuit{n}{m}$. The \emph{inputs}
(resp. \emph{outputs}) of $x$ are a finite function $ins$ of type $n
\rightarrow \data$ (resp. $outs$ of type $m \rightarrow \data$).  The
\emph{meaning} of $x$ is a relation $\invar{n}{m}{x}{ins}{outs}$
between $ins$ and $outs$ that we define by induction on $x$.
%
This is an abstract mathematical characterization, which may or may
not be computational (we will come back to this point later).

\subsubsection*{Abstracting from the implementation.} 
The meaning of a circuit is defined by induction on its structure:
this relation may be complex and may give informations about the
internal implementation of a circuit. Thus, we want to move from the
definition of this relation, for instance, to give high-level
specifications, or to abstract their behavior.
These abstractions can be expressed through the following kind of
entailment~\cite{melham}:
$$\forall ins, \forall outs,
\invar{}{}{\gate{RIPPLE~n}}{ins}{outs} \rightarrow R~outs~ins$$

We use \emph{data-abstraction}~\cite{melham} to be more
elegant. Indeed, a value of type
$\unit \oplus \sumn{}{n} \oplus \sumn{}{n} \rightarrow \bool$ is
isomorphic to a value of type $\bool \times \word{n} \times \word{n}$
(where $\word{n}$ is the type of integers from $0$ to $2^n$). We use
\emph{type-isomorphisms} to give tractable specifications for
circuits: we prove that the parametric $n$-bit adder depicted in
fig~\ref{fig:intro-ripple} implements the addition with carry function
on $\word{n}$.


    


\section{Formal development}\label{sec:definitions}
We now turn to define formally the concepts that were overviewed in
the previous section. We use Coq type-classes to structure our
development and parametrize code.

\subsection{Circuit interfaces}
We use arbitrary finite types (types with finitely many elements) as
interfaces for the circuits, i.e., as indexes for the wires. One can
create such finite types by using the disjoint-sum operator $\oplus$
and the one-element type $\unit$. This construction can be generalized
to $n$-ary disjoint sums written \coqe{sumn A n}, for a given
\coqe{A}.
However, using a single singleton type for all wires can be confusing:
there is no way to distinguish one $\unit$ from another, except by its
position in the type (which is frustrating). Hence, we use an infinite
family of singleton types $\unit_{x}$ where $x$ is a \emph{tag}. Circuits are parametrised by some tags, which
allows the Coq type-system to rule out some ill-formed
combinations. This tagging discipline allows to easily follow circuit
diagrams to define circuits in Coq, without much room for mistakes.

\begin{coq}
Inductive tag (t : string) : Type := _tag : tag t. (** we write $\unit_t$ for tag t*)  
\end{coq}

Finite types are defined as a class \coqe{Fin A} that packages a
duplicate-free list of all the elements of the type \coqe{A}, defined
along the lines of~\cite{ssreflect}.



\subsection{Type isomorphisms}
We use type-isomorphisms as ``lenses'' to express the specification of
circuits in user-friendly types, without loss of information.
In a nutshell, we define in Coq an isomorphism between two types $A$
and $B$ as a pair of functions $iso:A \rightarrow B$ and $uniso: A
\rightarrow B$ that are proved to be inverse of each other. We use the
notation $A \cong B$ for an isomorphism between $A$ and $B$, and we
define some notations for operations (or instances)that allow one to
build such isomorphisms in Fig.~\ref{fig:isi}.
The most important instance state the duality between disjoint-sums in
the domain of the finite functions to a cartesian product. 

\medskip

\begin{twolistings}
\begin{coq}
Class Iso (A B : Type) :={
 iso : A -> B;
 uniso : B -> A}.
\end{coq}
&
\begin{coq}
Class Iso_Props {A B: Type} (I : Iso A B):= {
 iso_uniso : forall (x : B), iso (uniso x) = x;
 uniso_iso : forall (x : A), uniso (iso x) = x}.
\end{coq}
\end{twolistings}

\newcommand\isosum{\bullet}
\newcommand\isounit[1]{\iota_{\tt #1}}
\newcommand\isoword[2]{\Phi_{#1}^{#2}}


\begin{figure}[b]
  \centering
\begin{mathpar}
  \inferrule*[left=$\cdot \isosum \cdot $]{ A \rightarrow \data \cong \sigma \and B
    \rightarrow \data \cong \tau } { A \oplus B \rightarrow \data
    \cong (\sigma \times \tau) } \and
 %
  %
  \inferrule*[left=$\isounit{x}$]{ }{ \unit_x \rightarrow \data \cong \data }
 \and
  \inferrule*[]{A \rightarrow \data \cong \sigma } { {\tt sumn}~A~n
    \rightarrow \data \cong {\tt vector}~\sigma~{\tt n} }
 \end{mathpar}  
  \caption{Isomorphisms between types}
  \label{fig:isi}
\end{figure}

\subsection{Plugs}
Rewiring circuits of type $\circuit{n}{m}$ are defined by mapping
output wires indexed by $m$ to input wires indexed by $n$. We define
plugs using usual Coq functions to get small and computational
definition of maps. (Note that, since we map the indexes of the wires,
there is no way to embed an arbitrary function inside our circuits to
compute, e.g., the addition of the value carried by two input wires.)

We give three examples: (a) is a circuit that forgets its first input
(types must be read bottom-up on diagrams); (b) is a circuit that
duplicates its inputs; (c) implements some re-ordering and duplication
of the wires. (We leave implicit the associativity of wires on the
diagrams.)

\begin{center}
  \begin{tabular}{ccc}
    (a) & (b) & (c) \\ 
    \includegraphics[width=0.2\linewidth, height=1cm]{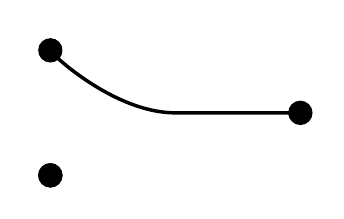}&
     \includegraphics[width=0.2\linewidth, height=1cm]{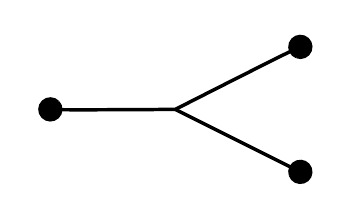}&
     \includegraphics[width=0.2\linewidth, height=1cm]{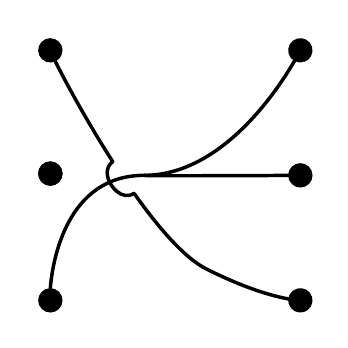}\\
     $\circuit{(n \oplus m)}{m}$ &
     $\circuit{n}{(n\oplus n)}$ &
     $\circuit{(n \oplus m \oplus p)}{(p \oplus (n \oplus n))}$ \\
  \end{tabular}
\end{center}

A possible implementation for (a) is \makebox{\coqe{fun x => inr x}}
and (b) can be implemented as
\makebox{\coqe{fun x => match x with inl e => e | inr e => e end}}.
If the type of the circuit gives enough informations, like the
examples above, it is possible to define such plugs using
proof-search. Indeed, plugs that deal with the associativity of the
wires, or even re-orderings, are completely defined by their type, and
we use tactics to write the map between wires (it amounts to some case
splitting and little automation).
Hence, in the formal definition of circuits, we omit the plugs that
deal with associativity or re-orderings of the wires, not only for the
sake of readability, but also because we do so in the actual Coq code:
we leave holes in the code (thanks to the Coq Program feature) that
will be filled automatically.

\subsection{Abstract syntax}\label{sec:syntax}
In the following, we use some basic gates from which all other
circuits are defined.
Hence, we parametrize the definition of circuits by the type of the 
 gates:
\begin{coq}
Class Techno := techno : Type -> Type -> Type. 
\end{coq}
The Fig.~\ref{fig:syntax} presents the dependent type that models
circuits, as defined in Coq.
This abstract syntax is strongly typed: it ensures that circuits built
using the provided combinators are well-formed: dimensions have to
agree, and it is not possible to connect circuits in the wrong
direction.  (Note that this is not anecdotal: if we were to describe
circuits with ports and wires, ensuring these properties would require
some boilerplate.)  
We denote serial composition (\coqe{Ser}) with the infix $\ser$
symbol, and parallel composition (\coqe{Par}) with $\para$. (Note that
these definitions do not deal with what transit in the wires.)

\begin{figure}[t]
  \centering
\begin{coq}
Context {tech : Techno}
Inductive $\mathbb{C}$  : Type -> Type -> Type :=
| Atom : forall {n m : Type} {Hfn : Fin n} {Hfm : Fin m}, techno n m -> $\mathbb{C}$ n m
| Plug : forall {n m : Type} {Hfn : Fin n} {Hfm : Fin m} (f : m -> n), $\mathbb{C}$ n m
| Ser : forall {n m p : Type}, $\mathbb{C}$ n m -> $\mathbb{C}$ m p -> $\mathbb{C}$ n p
| Par : forall {n m p q : Type}, $\mathbb{C}$ n p -> $\mathbb{C}$ m q ->  $\mathbb{C}$ (n \oplus m) (p \oplus q)
| Loop : forall {n m p : Type}, $\mathbb{C}$ (n \oplus p) (m \oplus p) -> $\mathbb{C}$ n m.
\end{coq}
\caption{Syntax}
\label{fig:syntax}
\end{figure}

\subsection{Structural specifications}
Let $\data$ be the type of what is carried in the wires. 
We now define the meaning relation for circuits. For a given circuit
of type $\circuit{n}{m}$, we build a relation between two functions of
type $n \rightarrow \data$ and $m \rightarrow \data$.
We define several operations on such functions, in order to express
the meaning relation in a legible manner:
\begin{coq}
Context {$\data$ : Type}.
Definition left  {n} {m} (x : (n \oplus m) -> $\data$) : n -> $\data$ := fun e => (x (inl _ e)).
Definition right {n} {m} (x : (n \oplus m) -> $\data$) : m -> $\data$ := fun e => (x (inr _ e)).
Definition lift {n} {m} (f : m -> n) (x : n -> $\data$) : m -> $\data$ := fun e => x (f e).
Definition app {n m} (x : n -> $\data$) (y : m -> $\data$) : n \oplus m -> $\data$ :=
  fun e => match e with inl e => x e | inr e => y e end. 
\end{coq}

We define the semantics of a given set of basic gates %
\coqe{tech: Techno}
by defining instances of the following type-class, (typically, one
instance for the boolean setting, and one instance in the stream of
boolean setting):
\begin{coq}
Class Technology_spec (tech : Techno) $\data$:= 
    spec : forall {a b: Type}, tech a b  -> (a -> $\data$)  -> (b -> $\data$) -> Prop.
\end{coq}

The meaning relation for circuits is generated by this parameter and
rules for each combinator. These rules are presented on
Fig.~\ref{fig:semantique} using inference rules rather than the
corresponding Coq inductive, for the sake of readability.

\newcommand{\dleft}[1]{{\tt left}~#1}
\newcommand{\dright}[1]{{\tt right}~#1} 
\newcommand{\dlift}[2]{{\tt lift}~#1~#2} 
\newcommand{\dapp}[2]{{\tt app}~#1~#2}

\begin{figure}[b]
  \centering
  \begin{mathpar}
    \inferrule*[left=KSer]{\invar{n}{m}{x}{ins}{middle} \and
      \invar{m}{p}{y}{middle}{outs}
    }
    { \invar{n}{p}{x \ser y}{ins}{outs} }
    \and
    \inferrule*[left=KPar]
    {\invar{n}{p}{x}{\dleft{ins}}{\dleft{outs}} \and
      \invar{m}{q}{y}{\dright{ins}}{\dright{outs}} 
    }
    { \invar{n\oplus m}{p \oplus q}{x \para y}{ins}{outs}}
    \\
    \inferrule*[left=KPlug]
    { }
    { \invar{n}{m}{{\tt Plug}~f}{ins}{\dlift{f}{ins}}}
    \and
    \inferrule*[left=KLoop]
    { \invar{n\oplus p}{m\oplus p}{x}{\dapp{ins}{r}}{\dapp{outs}{r}}}
    { \invar{n}{m}{{\tt Loop}~x}{ins}{outs}}

  \end{mathpar}
  \caption{Meaning of circuits (omitting the rule for Atom)}
  \label{fig:semantique}
\end{figure}

\subsection{Abstractions}
The meaning relation defines precisely the behavior of a circuit, but
cannot be used as it is. 
First, it may be too precise, e.g., with some internal details
leaking, or imposing constraints between the inputs and the outputs of
a circuit that are not relevant from an external point of view.
Second, it defines a constraint between the inputs and outputs of a
circuit as a relation between two functions $n \rightarrow \data$
and $m \rightarrow \data$, which is not user-friendly.
In his book~\cite{melham}, Melham defines two kinds of abstractions
that are relevant here:
behavioral abstraction (expressed through the logical entailment of a
weak specification $R$ by the meaning relation) 
and data-abstraction (when the specification is expressed in terms of
higher-level types than the above function types).

We combine these two notions to specify that a given circuit realises
a specification $R$ up-to two type isomorphisms, and to get more
concise specifications, we also define the fact that a circuit
implements a function \coqe{f} up-to isomorphisms:
\begin{coq}
Context {n m N M : Type} (Rn : (n->$\data$) $\cong$ N) (Rm : (m->$\data$) $\cong$ M). 
Class Realise (c : $\circuit{\tt n}{\tt m}$) (R : N -> M -> Prop) := realise: forall ins outs, 
\tab\tab$\invar{\tt n}{\tt m}{c}{\tt ins}{\tt outs}$ -> R (iso ins) (iso outs).
Class Implement (c : $\circuit{\tt n}{\tt m}$) (f : N -> M) := implement: forall ins outs,
\tab\tab$\invar{\tt n}{\tt m}{c}{\tt ins}{\tt outs}$ -> iso outs = f (iso ins).
\end{coq}

\subsection{Atoms and modular proofs}
We develop circuits in a modular way: to build a complex circuit, we
define a functor that takes as argument a module that packages the
implementations of the sub-components, and the proofs that they meet
some specification. %
This means that our proofs are hierarchical: we do not inspect the
definition of the sub-components when we prove a circuit.
These functors can then be applied to a module that contains a set of
basic doors (of type \coqe{Techno}) and its meaning relation (of type
\coqe{Technology_spec}).

\section{Proving some combinatorial circuits}\label{sec:comb}
In this section, we focus on acyclic combinational circuits, and
implements some arithmetic circuits. We assume a set of basic gates
(\gate{AND}, \gate{XOR} among others, that can all be defined and
proved correct starting from \gate{NOR} only). Wires carry booleans,
i.e., the meaning relation is defined on booleans for the basic
gates. \thomas{Check} We first illustrate our proof methodology on a
half-adder. Then, we present operations on $n$-bits integers, that
will be used to specify  $n$-bit adders.

\subsection{Proving a half-adder}
\begin{figure}[t]
  \centering
  \begin{tabular}{cc}
    \begin{minipage}[c]{0.4\linewidth}
      \begin{center}
        \includegraphics[width=0.5\linewidth]{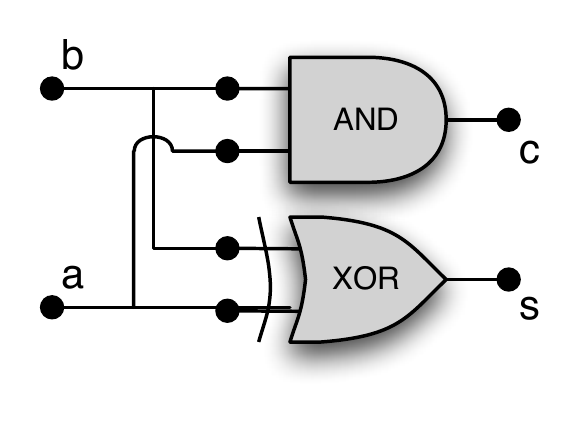}
      \end{center}
    \end{minipage}
    &
    \begin{coq}
Context a b s c : string. (*section variables*)
Definition HADD : $\mathbb{C}$ ($\unit_a \oplus \unit_b$) ($\unit_s \oplus \unit_c$) :=
Fork2 ($\unit_a \oplus \unit_b$) |> (XOR a b s  & AND a b c).  
    \end{coq}    
  \end{tabular}
  \caption{Definition of a half-adder} \label{fig:half-adder}
\end{figure}

A half-adder adds two $1$-bit binary numbers together, producing a
$1$-bit number and a carry out. However, they cannot be chained
together since they have no carry in input.  We present a diagram of
this circuit, along with its formal definition, in
Fig.~\ref{fig:half-adder}.  
The left-hand side of the following Coq excerpt is the statement we
prove: the circuit \gate{HADD} implements the function \coqe{hadd} on
booleans (defined as \coqe{\fun(a,b).(a \xorb b, a && b)}, where
$\oplus$ is the boolean exclusive-or, and $\wedge$ is the boolean and)
up-to isomorphisms (we use the notations from Fig.~\ref{fig:isi} for
isos). The Coq system ask us to give evidence of the right-hand side.

\medskip
\begin{twolistings}
\begin{coq}
Instance HADD_Spec : Implement 
($\isounit{a} \isosum \isounit{b}$) (* iso on inputs *)
($ \isounit{s} \isosum \isounit{c}$) (* iso on outputs *)
HADD hadd. 
\end{coq}
&
\begin{coq}
I : $\unit_{\tt a} \oplus \unit_{\tt b}$ -> \bool, O : $\unit_{\tt s} \oplus \unit_{\tt c}$ -> \bool
H : $\invar{\unit_{\tt a} \oplus \unit_{\tt b}}{\unit_{\tt s} \oplus \unit_{\tt c}}{\tt HADD}{\tt ins}{\tt outs}$
====================
@iso ($\isounit{s} \isosum \isounit{c}$) O = hadd (@iso ($\isounit{a} \isosum \isounit{b}$) I) 
\end{coq}
\end{twolistings}

\medskip

\noindent We have developed several tactics that help to prove this kind of
goals.
First, we automatically invert the derivation of the meaning relation
in the hypothesis \coqe{H}, following the structure of the circuit, to
get rid of parallel and serial combinators.  This leaves the user with
one meaning relation hypothesis per sub-component in the circuit
(plugs included).
Second, we use the type-class \coqe{Implement} as a dictionary of
interesting properties. This allows one to make fast-forward reasoning
by applying \coqe{implements} in any hypothesis stating a meaning
relation for a sub-component. The type-class resolution mechanism will
look for an instance of \coqe{Implement} for this sub-component, and
transform the ``meaning relation'' hypothesis into an equation. (Note
that at this point, the user may have to interact with the
proof-assistant, e.g., to choose other \coqe{Implement} instances than
the one that are picked automatically, but in many cases, this step is
automatic.) At this point, the goal looks like the left-hand side of
the following excerpt:

\medskip

\begin{twolistings}
\begin{coq}
I : $\unit_{\tt a} \oplus \unit_{\tt b}$ -> \bool, O : $\unit_{\tt s} \oplus \unit_{\tt c}$ -> \bool
M : $(\unit_{\tt a} \oplus \unit_{\tt b}) \oplus (\unit_{\tt a} \oplus \unit_{\tt b})$ -> \bool
H0: iso M = (fun x => (x,x)) (iso I)
H1: iso (left O) = uncurry \xorb (iso (left M))
H2: iso (right O)= uncurry \andb (iso (right M))
==========================
iso O = hadd (iso I)
\end{coq}
&
\begin{coq}
I: \bool * \bool, O: \bool * \bool,
M : (\bool * \bool) * (\bool * \bool), 
H0: M = (fun x => (x,x)) I
H1: fst O = uncurry \xorb (fst M)
H2: snd O = uncurry \andb (snd M)
==================
O = hadd I
\end{coq}
\end{twolistings}

\medskip

\noindent Third, we move to the right-hand side of the excerpt:
we massage the goal to make some \coqe{iso} commute with the
\coqe{left}, \coqe{right} and \coqe{app} operations, in order to
generalize the goal w.r.t. the isos. (Note that the user may be
required to interact with Coq if different isos are applied to
the same term in different equations.)
Finally, the proof context deals only with high-level data-types, and
functions operating on these.  The user may then prove the
``interesting'' part of the lemma.

\subsection{$n$-bits integers}\label{sec:nbit-integers}
From now, we use a dependently typed definition of $n$-bits integers,
along the lines of the fixed-size machine integers
of~\cite{Leroy-backend}. We omit the actual definitions of functions
when they can be infered from the type. In the
following, we prove that various (recursive) circuits implement the
\coqe{carry_add} function (that adds two $n$-bit numbers and a carry).

\begin{coq}
Record word (n:nat) := mk_word {val : Z; range: 0 $\le$ val < $2^n$}.   (* $\cword{n}$ *)

Definition repr n (x : Z) : $\word{n}$ := ...
Definition high n m (x : $\cword{(n+m)}$) : $\cword{m}$ := ...
Definition low n m (x : $\cword{(n+m)}$) : $\cword{n}$ := ... 
Definition combine n m (low : $\cword{n}$) (high : $\cword{m}$) : $\cword{(n+m)}$ := ...
Definition carry_add n (x y : $\cword{n}$) (b : \bool) : $\cword{n}$ * \bool := 
\tab\tablet e := val x + val y + (if b then 1 else 0) in (e mod $2^n$,$2^n$ $\le$ e)

Definition $\isoword{\tt x}{\tt n}$ : Iso ($\sumn{x}{n} \rightarrow \bool$) ($\word{n}$) := ...
\end{coq}

\subsection{Two specifications of a $1$-bit adder}
\begin{figure}[tb]
  \centering
  \begin{tabular}{cc}
    \begin{minipage}[c]{0.4\linewidth}
          \begin{center}
            \includegraphics[width=1\linewidth]{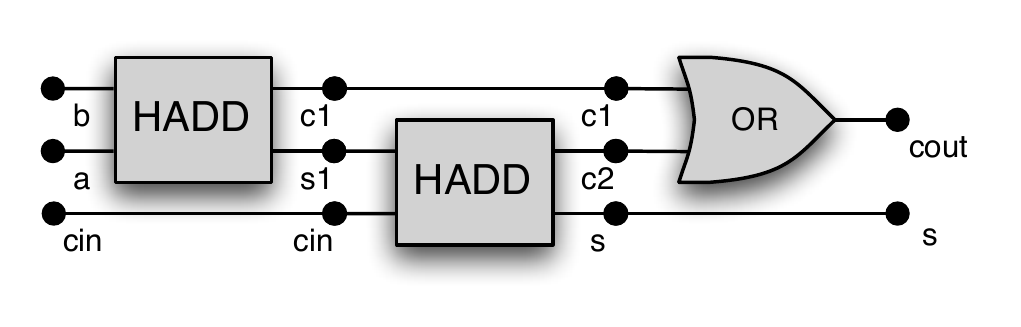}
      \end{center}
    \end{minipage}
    &
    \begin{coq}
Context a b cin sum cout : string. 
Program Definition FADD  :
 $\mathbb{C}$ ($\unit_{cin} \oplus (\unit_a \oplus \unit_b)$) ($\unit_{sum} \oplus \unit_{cout}$) :=
  (ONE $\unit_{cin}$ &  HADD a b "s1" "c1")
  |> ... (* associativity plug *)
  |> (HADD cin "s1" sum "c2" & ONE $\unit_{"c1"}$)
  |> ... (* associativity plug *)
  |> (ONE $\unit_{sum}$ & OR "c2" "c1" cout).
   \end{coq}
  \end{tabular}
  \caption{Definition of a full-adder}    \label{fig:full-adder} 
\end{figure}

A full-adder adds two $1$-bit binary numbers with a carry in,
producing a $1$-bit number and a carry out, and is built from two
half-adders. We present a diagram of this circuit, along with its
formal definition in Fig.~\ref{fig:full-adder}.
 
From this circuit, we can derive two specifications of interest.
First, the meaning of the full-adder can be expressed in terms of a
boolean function, that mimics the truth-table of the circuit.
Second, we can prove that this circuit actually implements the
\coqe{carry_add} function up-to isomorphism. 
Both these specifications are proved using the aforementioned tactics,
only the interesting parts differ.

\begin{twolistings}
\begin{coq}
Instance FADD_1 : Implement 
($ \isounit{cin} \isosum (\isounit{a} \isosum \isounit{b})$) (* iso on inputs *)
($ \isounit{sum} \isosum \isounit{cout}$) (* iso on outputs *)
FADD (fun (c,(x,y)) => 
\tab(x \oplus (y \oplus c),(x && y) || c && (x \oplus y))).
\end{coq}
&
\begin{coq}
Instance FADD_2 : Implement
($ \isounit{cin} \isosum (\isoword{a}{1} \isosum \isoword{b}{1})$) (* iso on inputs *)
($ \isoword{sum}{1} \isosum \isounit{cout}$) (* iso on outputs *)
FADD (fun (c,(x,y)) => 
\tabcarry_add 1 x y c). 
\end{coq}
\end{twolistings}

\thomas{Attention, FADD2, ce n'est pas les bons isis}

\subsection{ Ripple-carry adder}
\begin{figure}[t]
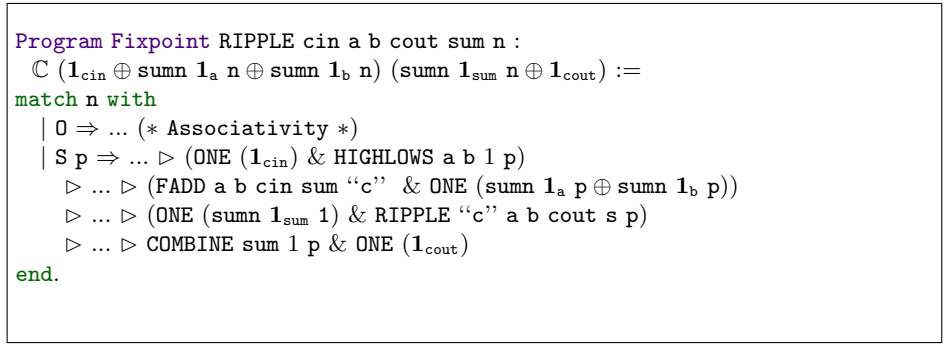

  \centering
  \begin{coq}
Program Fixpoint RIPPLE cin a b cout sum n :  
  $\circuit{(\unit_{\tt cin} \oplus \sumn{a}{n} \oplus \sumn{b}{n})}{(\sumn{sum}{n}\oplus \unit_{\tt cout})}$ := 
match n with 
\tab| O => ... (* Associativity *)
\tab| S p => ... |> (ONE ($\unit_{\tt cin}$) & HIGHLOWS a b 1 p)
\tab\tab|> ... |> (FADD a b cin sum ``c''  & ONE $(\sumn{a}{p} \oplus \sumn{b}{p})$)
\tab\tab|> ... |> (ONE ($\sumn{sum}{1}$) & RIPPLE ``c'' a b cout s p) 
\tab\tab|> ... |> COMBINE sum 1 p & ONE ($\unit_{\tt cout}$)
end.     
  \end{coq}
  \caption{Implementation of the ripple-carry-adder from Fig.~\ref{fig:intro-ripple}} \label{fig:ripple}
\end{figure}

We present in Fig.~\ref{fig:ripple} the formal definition of the ripple-carry
adder from Fig.~\ref{fig:intro-ripple} (we omit the rewiring plugs).
This definition is based on two new circuits to split wires, and
combine them. Indeed, to build a $1+n$-bit adder, the lowest-order
wire of each parameter is connected to a full-adder, while the $n$
high-order wires of each parameter are connected to another
ripple-carry adder. Conversely, the wires corresponding to the sum
must be combined together. %
We use two plugs to define the \gate{HL} and the \gate{COMBINE}
circuits.

\begin{coq}
Definition HL x n p : $\circuit{(\sumn{x}{(n + p)})}{(\sumn{x}{n} \oplus \sumn{x}{p})}$:= Plug ... 
Definition COMBINE x n p : $\circuit{(\sumn{x}{n} \oplus \sumn{x}{p})}{(\sumn{x}{(n + p)})}$:= Plug ...
\end{coq}

\noindent Then, we prove that these functions on wires implements
their counterparts on words. These gates are then easily combined
two-by-two to build \gate{HIGHLOWS} and \gate{COMBINES} that works
with two sets of wires at the same time to get more economical designs
(i.e., designs with less sub-components).

\medskip

\begin{twolistings}
\begin{coq}
Lemma HL_Spec x n p: Implement 
($\isoword{x}{n+p}$) ($\isoword{x}{n} \isosum \isoword{x}{p}$) (HL x n p)
(fun x => (low n p x, high n p x)).   
\end{coq}  
&
\begin{coq}
Lemma COMBINE_Spec x n p: Implement
($\isoword{x}{n} \isosum \isoword{x}{p}$) ($\isoword{x}{n+p}$) (COMBINE x n p)
(fun x => (combine n p (fst x) (snd x))). 
\end{coq}
\end{twolistings}

\medskip

\noindent Finally, we prove by induction on the size of the circuit
that it implements the high-level \coqe{carry_add} addition function
on words.  (Note that this is a high-level specification of the
circuit: the \coqe{carry_add} function is not recursive and disclose
nothing of the internal implementation of the device.)  This boils
down to the proof of lemma \coqe{add_parts}.

\begin{coq}
Lemma add_parts n m (xH yH: word m) (xL yL : word n) cin: 
\tablet (sumL,middle) := carry_add n xL yL cin in 
\tablet (sumH,cout) := carry_add m xH yH middle in 
\tablet sum := combine n m sumL sumH in 
\tabcarry_add (n + m) (combine n m xL xH)(combine n m yL yH) cin = (sum,cout).

Instance RIPPLE_Spec cin a b cout sum n : Implement (RIPPLE cin a b cout s n)
\tab($\isounit{cin} \isosum (\isoword{a}{n} \isosum \isoword{b}{n})$)\tab($\isoword{sum}{n} \isosum \isounit{cout}$)\tab(fun (c,(x,y)) => carry_add c x y).
\end{coq}

This design is simple (a linear chain of $1$-bit adders) and slow
(each full-adder must wait for the carry-in bit from the previous
full-adders).
In the next subsection, we address the case of a more efficient adder,
which is incidentally more complicated, and a better benchmark for our
methodology.

\subsection{Divide and conquer adder}\label{sec:dc-adder}
A text-book~\cite{aho-ullman} solution to improve on the delay of the
ripple-carry adder is to use a divide and conquer scheme, and to
compute both the sum when there is a carry in, and the sum when there
is no carry in. It is then possible to compute at the same time the
sum for the high-order bits, and the sum for the low order bits.
Hence, we build a circuit that computes four pieces of data: $s$
(resp. $t$), the $n$-bit sum of the inputs, assuming that there is no
carry in (resp. assuming that there is a carry in); $p$ the
\emph{carry-propagate} bit (resp. $g$ the \emph{carry-generate} bit),
which is true when there is a carry out of the circuit, assuming that
there is a carry in (resp. that there is no carry in).

\begin{figure}[tb]
  \centering
    \begin{tabular}{cc}
    \begin{minipage}[c]{0.4\linewidth}
          \begin{center}
            \includegraphics[width=1\linewidth]{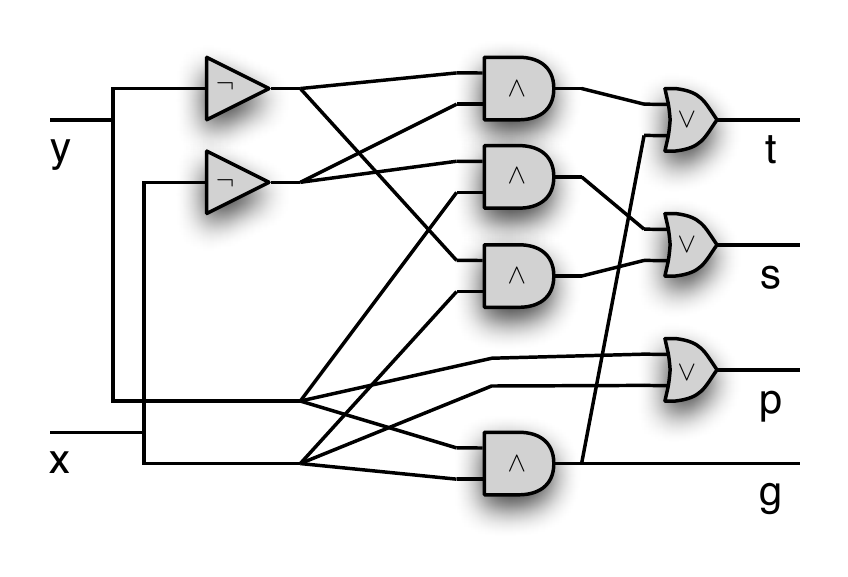}
      \end{center}
    \end{minipage}      
    &  
    \begin{minipage}[c]{0.4\linewidth}
          \begin{center}
            \includegraphics[width=1\linewidth]{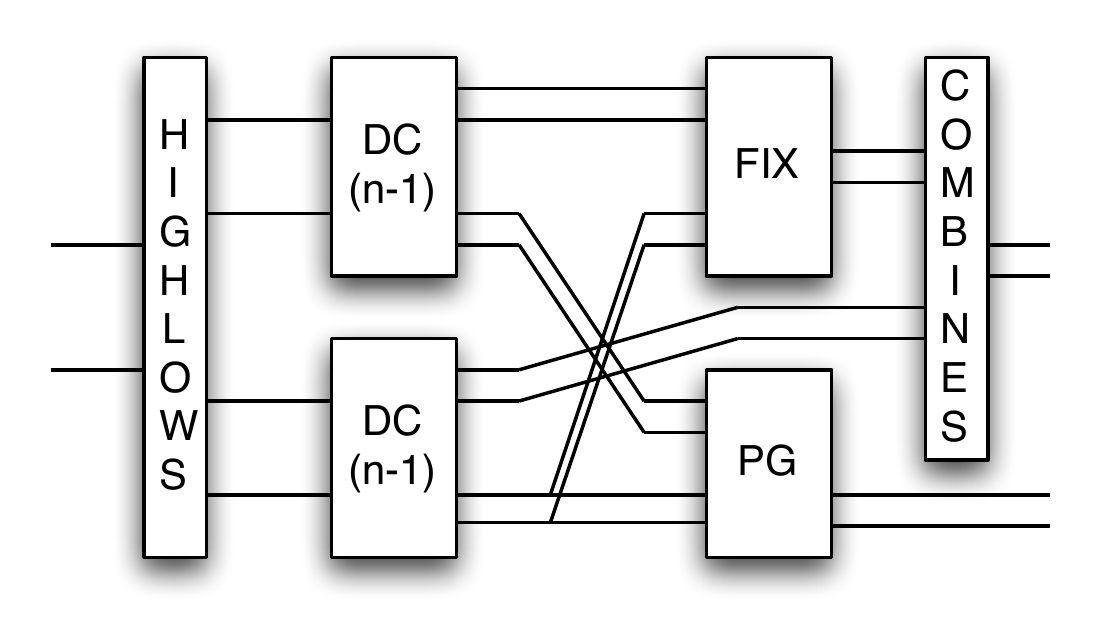}
      \end{center}
    \end{minipage}
    
  \end{tabular}
  \caption{Divide and conquer adder}
  \label{fig:dc}
\end{figure}

We provide a diagram in Fig.~\ref{fig:dc} that depicts the base case
and the recursive case, but we omit the actual Coq implementation, for
the sake of readability.
We prove that this circuit implements the following Coq function:
\begin{coq}
Definition dc n :$\word{2^n}$ * $\word{2^n}$  -> \bool * \bool * $\word{2^n}$ * $\word{2^n}$ := fun (x,y) => 
let (s,g) := carry_add $2^n$ x y false in 
let (t,p) := carry_add $2^n$ x y true in (g,p,s,t).      
\end{coq}

Again, this is a high-level specification w.r.t. the definition of the
circuit: it does not disclose how the circuit compute its results (for
instance, the \coqe{dc} function is not recursive).
In a nutshell, the circuit computes in parallel the 4-uple of results
for the high-order and low-order part of the inputs.
Then, the propagate and generate bits for both parts can be combined
by the \gate{PG} circuit to compute the propagate and generate bits
for the entire circuit.  
%
In parallel, the \gate{FIX} circuit is made of two $2^{n-1}$-bit
multiplexers (easily defined with a fixpoint using $1$-bit
multiplexers), and update the high-order parts of the sum, w.r.t. the
propagate and generate carry-bits of the low-order adder.

\section{Sequential circuits: time and loops}\label{sec:time}
While we have focused our case studies on combinational circuits, our
methodology can be applied to sequential circuits, with or without the
loops that were allowed in the syntax of circuits in
\S\ref{sec:syntax}.  In this section, the wires carry streams of
booleans (of type \coqe{nat -> \bool}), and we assume a basic gate
\gate{DFF} that implements the \coqe{pre} function (in the particular
case of booleans):

\begin{twolistings}
\begin{coq}
Definition pre {A} (d : A): 
stream A -> stream A := fun f t => 
match t with | 0 => d | S p => f p end.
\end{coq}
&
\begin{coq}
Instance DFF_Realise_stream {a  out}: 
Implement  (DFF a out) ($\isounit{a}$) ($\isounit{out}$) 
(pre false).
\end{coq}
\end{twolistings}

\subsubsection*{A buffer.}
A \gate{DFF} delays one wire by one unit of time; a \gate{FIFO} buffer
generalize this behavior in two dimensions, by chaining layers of
\gate{DFF} one after another. This circuit is simple, but is a good example for the use of high-level combinators. 

These combinators capture the underlying regularity in some common
circuit pattern, for instance replicating a sub-component in a serial
or parallel manner.

\medskip
\begin{twolistings}
\begin{coq}
Variable CELL : $\circuit{n}{n}$. 
Fixpoint COMPOSEN k : $\circuit{n}{n}$ :=
match k with 
| 0 => Plug id 
| S p => CELL |> (COMPOSEN p)    
end.  
\end{coq}
&
\begin{coq}
Variable CELL : $\circuit{n}{m}$.
Fixpoint MAP k : $\circuit{(\sumn{n}{k})}{(\sumn{m}{k})}$:=
match k with 
| 0 => Plug id 
| S p => CELL & (MAP p)
end.      
\end{coq}  
\end{twolistings}

\medskip

We prove that the \coqe{COMPOSEN} combinator implements a higher-order
iteration function, up-to isomorphism: if \coqe{CELL} implements a
given function \coqe{f}, then \coqe{COMPOSEN k} implements the
iteration of \coqe{f}.  Respectively, we prove that the \coqe{MAP}
circuit implements the higher-order \coqe{map} function on vectors.
Hence, a \coqe{FIFO} buffer in one-line, and we prove
that it implements the function below.

\begin{coq}
Definition FIFO x n k : $\circuit{(\sumn{x}{k})}{(\sumn{x}{k})}$ := COMBINEN (MAP (DFF x x) k) n. 
Definition fifo n k (v : stream (vector \bool k)) : stream (vector \bool k) :=
\tabfun t => if n < t then v (t - n) else Vector.repeat k false. 
Remark useful_iso :  ${\tt sumn}~\unit~{\tt n} \rightarrow {\tt stream}~\bool \cong  {\tt stream}~({\tt vector}~\bool~{\tt n})$ := ...
\end{coq}

The proof of this specification relies on the above useful isomorphism
between groups of wires that carries streams of booleans, and a stream
of vectors of booleans. The proof that the circuit implements a
function on streams is done in the same fashion as the proofs from the
previous section.

\subsubsection*{A memory element.} 
\begin{figure}[tb]
  \centering
  \begin{tabular}{cc} 
\begin{minipage}{0.55\linewidth}
\begin{coq}
Context a load out : string.
Program Definition REGISTER: 
$\circuit{(\unit_{\tt load} \oplus \unit_{\tt a})}{\unit_{\tt out}}$ :=
@Loop $(\unit_{\tt load} \oplus \unit_{\tt a})$ $\unit_{\tt out}$ $\unit_{\tt out}$ 
(... |> MUX2 a out load "in_dff" 
|> DFF "in_dff" out |> Fork2 $\unit_{\tt out}$).
\end{coq}   
\end{minipage}
&
\begin{minipage}{0.4\linewidth}
  \includegraphics[height=3cm]{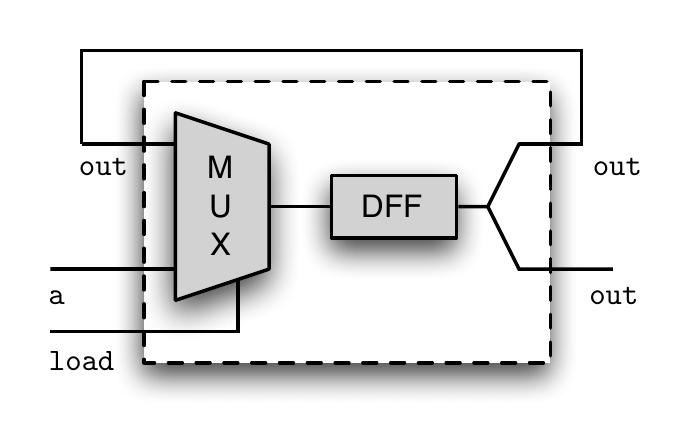}
\end{minipage}
 \end{tabular}
  \caption{A memory element}
  \label{fig:register}
\end{figure}

Our next goal is to demonstrate how
we deal with state-holding structures. Hence, we turn to the
implementation of a $1$-bit memory element, as implemented in
Fig.~\ref{fig:register}. 
The register is meant to hold $1$-bit of information through time,
which does not fit nicely in the \coqe{Implement} framework (we
cannot easily express the meaning of the register in terms of a stream
transformer). Hence, we use a relational specification through the
use of \coqe{Realise}:

\newcommand\stream[1]{{\tt stream}~{#1}}
\begin{coq}
Instance Register_Spec : Realise
(... :  $\unit_{\tt load} \oplus \unit_{\tt a} \rightarrow \stream{\bool}
\cong
\stream{\bool*\bool}$) ($\isounit{out}$) REGISTER 
(fun (ins : stream (\bool * \bool)) (outs : stream \bool) => 
\tabouts = pre false (fun t => if  fst (ins t) then snd (ins t) else outs t)).
\end{coq}

Here, the state of the register is stored inside the history of the
stream (the previous values that were taken by the output). While we
do not advocate that this is the nicest way to reason about state
holding devices, we were able to prove this specification in the same
fashion as the previous combinatorial devices. We leave more thorough
investigation of state-holding devices to future work.

\section{Interesting corollaries}\label{sec:free}
We now turn on to investigate some interesting consequences of the use
of a concrete data-type to represent circuit. 
First, we prove that the behavior of combinatorial circuits without
delay can be lifted to the stream setting.  Second, we build some
functions (or interpretations~\cite{Bjesse98lava:hardware}) that
operates on circuits.
\subsubsection{Lifting combinatorial circuits.} The meaning relation
is parametrized by the semantics of the basic gates. This can be put
to good use to prove the functional correction of some designs in the
boolean setting, and then, to mechanically lift this proof of
functional correction to the boolean stream setting (for the same set
of gates). For instance, if a loop-less and delay-less circuit
implements a function \coqe{f} in the boolean setting, we can prove
that the very same circuit implements the function \coqe{Stream.map f}
in the boolean stream setting.  

\subsubsection*{Simulating and checking designs. }
One key feature of our first-order encoding of circuits in Coq is that
it allows to check designs by simulation before attempting to prove
them. %
This verification is done on the same definition than the one which
will be proved later, allowing a seamless approach. While simulation
cannot be done on circuits parametrized by a size, this remains a
valuable help to avoid dead-ends.  We define a simulation function \coqe{sim} that works
on loop-free circuits, if the user provide a computational
interpretation of each basic gate. For instance, it allows to simulate
the adders of \S\ref{sec:comb}.


\subsubsection{Delay and pretty-printing.} Using the same ideas, we
can build functions that compute the list of gates of circuits (with
or without loops), or compute the length of the critical path in
combinatorial circuits. While this is more anecdotal, and less
directly useful than the previous simulation function, these functions
are still interesting: one could, for instance, prove that some
complex designs meet some time (or gate-count) complexity
properties. (Note that is the only place where we exploit the
finiteness of types.)

\section{Comparisons with related works}\label{sec:related-works}
\subsubsection{Verifying circuits with theorem provers.} 
There has been a substantial amount of work on specification and
verification of hardware in HOL. In \cite{UCAM-CL-TR-77,melham}, HOL
is used as a hardware description language and as a formalism to prove
that a design meets its specification. They model circuits as
predicates in the logic, using a shallow-embedding that merges the
architecture of a circuit and its behavior.
Building on the former methodology,
\cite{DBLP:journals/fac/SlindOIG07} defines a compiler from a
synthetisable subset of HOL that creates correct-by-construction
clocked synchronous hardware implementations of mathematical functions
specified in HOL.
This methodology allows the designer to focus on high-level
abstraction instead of reasoning and verifying at the gate level,
admitting the existence of some base low-level circuits (like the
addition on words ~\cite{UCAM-CL-TR-682}).
By contrast, our work complement their behavioral ``correct by
design'' synthesis from a subset of the high-level language of the
theorem-prover with structural verification of circuits.

In the Boyer-Moore theorem prover (untyped, quantifier-free and
first-order), Brock and Hunt proved the correctness of functions that
generate correct hardware designs. They studied the correctness of an
arithmetic and logic unit, parametrised by a
size~\cite{DBLP:conf/msiw/HuntB89}. This verified synthesis approach
was used to verify a microprocessor
design~\cite{DBLP:journals/fmsd/BrockH97}. While our proofs are not as
automated, and our examples are less ambitious, we are able to prove
higher-order circuits. Moreover, the dependent-types we use are
helpful when defining complex circuits.

In Coq, Paulin-Mohring~\cite{paulin-mohring-multiplier} proved the
correction of a multiplier unit, using a shallow-embedding similar to
the methodology used in HOL: circuits are modelled as functions of the
Coq language.
More recently, \cite{certifying-circuits-in-type-theory} investigated
how to take advantage of dependent types and co-inductive types in
hardware verification: they use a shallow embedding of Mealy automata
to describe sequential circuits.
By contrast with both works, we use a deep-embedding of circuits in
Coq, that makes explicit the definition of circuits. We still need to
investigate the examples of sequential circuits studied in these
papers.  

\subsubsection{Algebraic definitions of circuits.}
Circuit diagrams have nice algebraic properties. Lafont~\cite{lafont}
studied the algebraic theory of boolean circuits and
Hinze~\cite{algebra-scans} studied the algebra of parallel prefix
circuits. Both settings are close to ours: however the former focused
on the algebraic structure of circuits, while the latter defined
combinators that allows to model (and prove correct using algebraic
reasoning) all standard designs of a restricted class of circuits.

\subsubsection{Functional languages in hardware design.} Sheeran
\cite{DBLP:journals/jucs/Sheeran05} made a thorough review of the use
of functional languages in hardware design, and of the challenges to
address. Our work is a step toward one of them: the design and
verification of parametrized designs, through the use of circuit
combinators. %
Lava~\cite{Bjesse98lava:hardware} is a language embedded in Haskell to
describe circuits, allowing one to define parametric circuits or
higher-order combinators. While much of our goals are common, one key
difference is that our encoding of circuits in Coq avoid the use of
bound variables (we use only combinators). Moreover, we use dependent
types, that are required to deal precisely with parametric
circuits. Finally, we prove the correctness of these parametric
circuits in Coq, while verification in Lava is reduced to the
verification of finite-size circuits.

\section{Conclusion and future works}
We have presented a deep-embeding of circuits in the Coq
proof-assistant that allows to build and reason about circuits,
proving high-level specifications through the use of type-isomorphism.
We have demonstrated that dependent types are useful to prove
automatically some well-formedness conditions on the circuits, and
helps to avoid time consuming mistakes.
Then, we proved by induction the correctness of some arithmetic
circuits of parametric size: this could not have been possible without
mimicking the structure of the usual circuit diagrams to define
circuit generators in Coq. 
The formal development accompanying this paper is available from the
authors web-page \cite{coquet}.

In the immediate future, we plan to continue the case studies
described in \S\ref{sec:comb}. In particular, we would like to
investigate how to construct parallel prefix circuits in our
framework~\cite{algebra-scans,DBLP:journals/jucs/Sheeran05}, and to
investigate combinational multipliers.
In the more distant future, it would be interesting to study some
front-ends to automatically generate some circuits: this could range
to the reduction of the boiler-plate inherent to the definition of
plugs, to the compilation of circuits from automaton. A major
inspiration on behavioral synthesis is the work of Ghica~\cite{GoS}.
We also look forward to study how our methodology applies to other
settings that booleans or streams of booleans. For instance, if we
move from booleans to the three-valued Scott's domain (unknown, true,
false), we may interpret circuits in the so-called constructive
semantics. We also hope that some of our methods could be applied to
the probabilistic setting.

\bibliography{bib}

\begin{thebibliography}{10}

\bibitem{aho-ullman}
A.~V. Aho and J.~D. Ullman.
\newblock {\em Foundations of {C}omputer {S}cience}.
\newblock Computer Science Press, W. H. Freeman and Company, 1992.

\bibitem{Bjesse98lava:hardware}
P.~Bjesse, K.~Claessen, M.~Sheeran, and S.~Singh.
\newblock {L}ava: {H}ardware {D}esign in {H}askell.
\newblock In {\em Proc.\ ICFP}, pages 174--184. ACM Press, 1998.

\bibitem{coquet}
T.~Braibant.
\newblock Coquet: a coq library for verifying hardware.
\newblock \url{http://sardes.inrialpes.fr/~braibant/coquet}, June 2011.

\bibitem{DBLP:journals/fmsd/BrockH97}
B.~Brock and W.~A.~Hunt Jr.
\newblock {The DUAL-EVAL Hardware Description Language and Its Use in the
  Formal Specification and Verification of the FM9001 Microprocessor}.
\newblock {\em {Formal Methods in System Design}}, 11(1):71--104, 1997.

\bibitem{brown-hutton}
C.~Brown and G.~Hutton.
\newblock Categories, allegories and circuit design.
\newblock In {\em Proc.\ LICS}, pages 372--381. IEEE Computer Society, 1994.

\bibitem{certifying-circuits-in-type-theory}
S.~Coupet-Grimal and L.~Jakubiec.
\newblock Certifying circuits in type theory.
\newblock {\em Formal Asp. Comput.}, 16(4):352--373, 2004.

\bibitem{GoS}
D.~R. Ghica.
\newblock {Geometry of synthesis: a structured approach to VLSI design}.
\newblock In {\em Proc.\ POPL}, pages 363--375, 2007.

\bibitem{ssreflect}
G.~Gonthier and A.~Mahboubi.
\newblock {An introduction to small scale reflection in Coq}.
\newblock {\em {Journal of Formalized Reasoning}}, 3(2):95--152, 2010.

\bibitem{UCAM-CL-TR-77}
M.~Gordon.
\newblock {Why higher-order logic is a good formalisation for specifying and
  verifying hardware}.
\newblock Technical Report UCAM-CL-TR-77, Cambridge Univ., Computer Lab, 1985.

\bibitem{hanna-veritas}
F.~K. Hanna, N.~Daeche, and M.~Longley.
\newblock Veritas$^{\mbox{+}}$: A specification language based on type theory.
\newblock In {\em Hardware Specification, Verification and Synthesis}, LNCS,
  pages 358--379. Springer, 1989.

\bibitem{harrison-euclidean}
J.~R. Harrison.
\newblock A {HOL} theory of {E}uclidean space.
\newblock In J.~Hurd and T.~Melham, editors, {\em Proc.\ TPHOLs 2005}, volume
  3603 of {\em LNCS}, pages 114--129. Springer, 2005.

\bibitem{algebra-scans}
R.~Hinze.
\newblock An {A}lgebra of {S}cans.
\newblock In Dexter Kozen and Carron Shankland, editors, {\em MPC}, LNCS, pages
  186--210. Springer, 2004.

\bibitem{UCAM-CL-TR-682}
J.~Iyoda.
\newblock {Translating HOL functions to hardware}.
\newblock Technical Report UCAM-CL-TR-682, University of Cambridge, Computer
  Laboratory, April 2007.

\bibitem{DBLP:conf/msiw/HuntB89}
W.~A.~Hunt Jr. and B.~Brock.
\newblock {The Verification of a Bit-slice ALU}.
\newblock In Miriam Leeser and Geoffrey Brown, editors, {\em {Hardware
  Specification, Verification and Synthesis}}, volume 408 of {\em LNCS}, pages
  282--306. Springer, 1989.

\bibitem{lafont}
Y.~Lafont.
\newblock Towards an algebraic theory of boolean circuits.
\newblock {\em Journal of Pure and Applied Algebra}, 184:2003, 2003.

\bibitem{Leroy-backend}
X.~Leroy.
\newblock A formally verified compiler back-end.
\newblock {\em Journal of Automated Reasoning}, 43(4):363--446, 2009.

\bibitem{melham}
T.~Melham.
\newblock {\em Higher Order Logic and Hardware Verification}, volume~31 of {\em
  Cambridge Tracts in Theoretical Computer Science}.
\newblock Cambridge University Press, 1993.

\bibitem{paulin-mohring-multiplier}
C.~Paulin-Mohring.
\newblock Circuits as {S}treams in {C}oq: {V}erification of a {S}equential
  {M}ultiplier.
\newblock In {\em TYPES}, pages 216--230, 1995.

\bibitem{sheeran-muFP}
M.~Sheeran.
\newblock {$\mu$FP, A Language for VLSI Design}.
\newblock In {\em {LISP and Functional Programming}}, pages 104--112, 1984.

\bibitem{DBLP:journals/jucs/Sheeran05}
M.~Sheeran.
\newblock {H}ardware {D}esign and {F}unctional {P}rogramming: a {P}erfect
  {M}atch.
\newblock {\em J. UCS}, 11(7):1135--1158, 2005.

\bibitem{DBLP:journals/fac/SlindOIG07}
K.~Slind, S.~Owens, J.~Iyoda, and M.~Gordon.
\newblock Proof producing synthesis of arithmetic and cryptographic hardware.
\newblock {\em Formal Asp. Comput.}, 19(3):343--362, 2007.

\end{thebibliography}
\end{document}